\documentclass[aps,prd,twocolumn,superscriptaddress,nofootinbib]{revtex4}

\usepackage{hyperref}
\usepackage{amsmath,amssymb,amsfonts}
\usepackage{color}
\usepackage{graphicx}
\usepackage{enumerate} % advanced enumerate environment
\usepackage{colordvi} % for color text
\usepackage{bm}% bold math
\usepackage{multirow}

\newcommand{\mpcoh}{\,h^{-1}\,{\rm Mpc}}

\newcommand{\bfk}{\boldsymbol{k}}
\newcommand{\bfx}{\boldsymbol{x}}

\newcommand{\bfr}{\boldsymbol{r}}

\newcommand{\bfq}{\boldsymbol{q}}

\newcommand{\xigg}{\xi^{\rm gg}}
\newcommand{\xigp}{\xi^{{\rm g}+}}
\newcommand{\xip}{\xi^{+}}
\newcommand{\xim}{\xi^{-}}

\newcommand{\xiX}{\xi^{\rm X}}

\newcommand{\PX}{P^{\rm X}}

\newcommand{\PhX}{\hat{P}^{\rm X}}
\newcommand{\Phgg}{\hat{P}^{\rm gg}}
\newcommand{\PhgE}{\hat{P}^{\rm gE}}
\newcommand{\PhEE}{\hat{P}^{\rm EE}}

\newcommand{\Pgglobs}{P^{\rm gg,obs}_{\ell}}
\newcommand{\PtXlmobs}{\wt{P}^{\rm X,obs}_{\ell,m}}

\newcommand{\PtgElmobs}{\wt{P}^{\rm gE,obs}_{\ell,m}}
\newcommand{\PtEElmobs}{\wt{P}^{\rm EE,obs}_{\ell,m}}

\newcommand{\PXobs}{P^{\rm X,obs}}

\newcommand{\xiggl}{\xi^{\rm gg}_{\ell}}

\newcommand{\xitgp}{\wt{\xi}^{{\rm g}+}}
\newcommand{\xitm} {\wt{\xi}^{-}}

\newcommand{\xitgplm}{\wt{\xi}^{{\rm g}+}_{\ell,m}}
\newcommand{\xitmlm}{\wt{\xi}^{-}_{\ell,m}}
\newcommand{\xitXlm}{\wt{\xi}^{X}_{\ell,m}}

\newcommand{\xigglobs}{\xi^{\rm gg,obs}_\ell}
\newcommand{\xitgplmobs}{\wt{\xi}^{{\rm g}+,{\rm obs}}_{\ell,m}}
\newcommand{\xitmlmobs}{\wt{\xi}^{-,{\rm obs}}_{\ell,m}}

\newcommand{\Pdd}{P_{\delta\delta}}
\newcommand{\Pdt}{P_{\delta\Theta}}
\newcommand{\Ptt}{P_{\Theta\Theta}}

\newcommand{\gammaE}{\gamma_{\rm E}}
\newcommand{\gammaB}{\gamma_{\rm B}}
\newcommand{\sigv}{\sigma_{\rm v}}

\newcommand{\WX}{W_{\rm X}}

\newcommand{\Wgp}{W_{{\rm g}+}}

\newcommand{\Wpm}{W_{\pm}}

\newcommand{\bK}{b_{\rm K}}
\newcommand{\deltag}{\delta_{\rm g}}

\newcommand{\rmin}{r_{\rm min}}
\newcommand{\rmax}{r_{\rm max}}

\newcommand{\DA}{D_{\rm A}}
\newcommand{\DAf}{D_{\rm A}^{\rm fid}}
\newcommand{\Hf}{H^{\rm fid}}

\newcommand{\hDA}{\alpha_\perp}
\newcommand{\hH}{\alpha_\parallel}

\newcommand{\be}{\begin{equation}}
\newcommand{\ee}{\end{equation}}

\newcommand{\wt}{\widetilde}
\newcommand{\bftheta}{{\boldsymbol \theta}}

\newcommand{\asiaa}{Academia Sinica Institute of Astronomy and Astrophysics (ASIAA), No. 1, Section 4, Roosevelt Road, Taipei 106319, Taiwan}
\newcommand{\ipmu}{Kavli IPMU (WPI), UTIAS, The University of Tokyo, Kashiwa, Chiba 277-8583, Japan}

\begin{document}

\title{
Joint Geometric and Dynamical Constraints on Cosmology from \\ Anisotropies in Galaxy Intrinsic-Alignment Correlations
}

\author{Teppei Okumura}
\email{tokumura@asiaa.sinica.edu.tw}
\affiliation{\asiaa}
\affiliation{\ipmu}

\date{\today} 

\begin{abstract}
We present the first joint cosmological analysis to extract both geometric and dynamical information from galaxy intrinsic alignments (IAs). 
Using BOSS spectroscopy cross-matched with galaxy shape measurements from the DESI Legacy Imaging Surveys, we measure anisotropic galaxy density--intrinsic ellipticity (GI) and intrinsic ellipticity (II) correlations over $0.43\leq z\leq0.7$ 
and decompose their spin-dependent angular structure into associated Legendre multipoles. 
The measured GI correlation exhibits the expected baryon acoustic oscillation (BAO) structure, while its anisotropy provides geometric information complementary to galaxy clustering. 
By jointly modeling redshift-space and Alcock-Paczynski distortions, we constrain the growth rate parameter $f\sigma_8$, the angular-diameter distance $\DA$, and the Hubble expansion rate $H$. Relative to galaxy clustering alone, adding IA reduces their fractional uncertainties by 32\%, 18\%, and 29\%, respectively. 
By mapping these constraints onto a flat $w_0$CDM model, IA also tightens the constraints on $w_0$, $\Omega_m$, and $H_0$;
however, the $w_0$ constraint is sensitive to the minimum scale included in the analysis.
Our results establish anisotropic galaxy shapes as an additional source of geometric and dynamical information for spectroscopic cosmology.
\end{abstract}

\maketitle 

%%%%%%%%%%%%%%%%%%%%%%%%%%%%%%%%%%%%%%%%%
% Section 1
%%%%%%%%%%%%%%%%%%%%%%%%%%%%%%%%%%%%%%%%%

%\section{Introduction} \label{sec:introduction}

%{\it Introduction}---
Observations of the large-scale structure (LSS) of the universe with galaxy redshift surveys provide a powerful means of constraining cosmological models \cite{Peebles:1980,Weinberg:2013}.
Galaxy surveys probe the origin of cosmic acceleration through geometric distortions measured with baryon acoustic oscillations (BAO) \cite{Peebles:1970,Sunyaev:1970} and dynamical distortions measured with redshift-space distortions (RSD) \cite{Kaiser:1987}
\cite[e.g.,][]{Peacock:2001,Tegmark:2004,Eisenstein:2005,Cole:2005,Okumura:2008,Guzzo:2008,Okumura:2016,Alam:2017}.
Recent analyses by the Dark Energy Spectroscopic Instrument (DESI) \cite{DESI_Collaboration:2016}, when combined with cosmic microwave background and Type Ia supernova data, have reported a statistical preference for dynamical dark energy over a cosmological constant \cite{DESI:2025a}.
Although $\Lambda$CDM remains consistent with the data in broader parameter spaces \cite{Roy_Choudhury:2024}, upcoming surveys such as the Subaru Prime Focus Spectrograph \cite{Takada:2014} and \textit{Euclid} \cite{Laureijs:2011} will test this result with substantially larger samples.

As observations improve, it becomes increasingly important to extract all available cosmological information from the LSS.
Galaxy intrinsic alignment (IA), long studied as a contaminant of gravitational-lensing measurements \citep{Heavens:2000,Croft:2000,Hirata:2004,Mandelbaum:2006,Okumura:2009,Okumura:2009a,Joachimi:2011,Singh:2015,Singh:2016,Tonegawa:2022,Hervas_Peters:2025}, has emerged as a complementary cosmological probe.
Three-dimensional ellipticity correlations are sensitive to RSD and BAO, providing dynamical and geometric information complementary to galaxy clustering \cite{Chisari:2013,Okumura:2017a,Okumura:2019,Taruya:2020,Okumura:2022,Inoue:2025}.

The first cosmological constraint from galaxy IA was obtained from Sloan Digital Sky Survey (SDSS) and SDSS-III Baryon Oscillation Spectroscopic Survey (BOSS) data \cite{Okumura:2023}, building on the theoretical models developed in Refs.~
\cite{Okumura:2020,Okumura:2020a,Okumura:2024}.
That analysis did not detect a BAO feature in the IA statistics and obtained only an RSD constraint on the growth rate.
Ref.~\cite{Xu:2023} 
subsequently detected isotropic BAO in IA statistics, but did not measure the anisotropy required to separate radial and transverse distance information through the Alcock--Paczynski (AP) effect \cite{Alcock:1979}, no geometric cosmological constraint from IA had yet been obtained.
A hint of the BAO feature in the angular correlation function of galaxy shapes in photometric surveys was reported by Ref.~\cite{Chan:2026}.

In this paper, we take the next step by exploiting the anisotropy of IA correlations. 
We measure the galaxy density--intrinsic ellipticity (GI) and intrinsic 
ellipticity--ellipticity (II) correlations of BOSS galaxies using galaxy shapes 
measured from DESI imaging, and decompose their spin-dependent angular structure 
into associated Legendre multipoles. This enables us, for the first time, to use 
IA anisotropies not only to probe the growth of structure through RSD but also to 
constrain cosmic geometry through AP distortions. In a joint analysis with galaxy 
clustering, we constrain the growth rate $f(z)\sigma_8(z)$, the 
angular-diameter distance $\DA(z)$, and the Hubble expansion rate $H(z)$. We then 
map these constraints onto a spatially flat $w_0$CDM model to constrain the 
dark-energy equation-of-state parameter $w_0$, the matter density parameter 
$\Omega_m$, and the Hubble constant $H_0$.
Throughout this paper, we adopt the flat $\Lambda$CDM model determined by Ref.~\cite{Planck-Collaboration:2020} as the fiducial cosmology.

%%%%%%%%%%%%%%%%%%%%%%%%%%%%%%%%%%%%%%%%%
% Section 2
%%%%%%%%%%%%%%%%%%%%%%%%%%%%%%%%%%%%%%%%%

%\section{SDSS Galaxy Samples}\label{sec:sdss}

{\it Galaxy density and shape samples}---We analyze the galaxy distribution over $0.43 \leq z \leq 0.70$ using the publicly available SDSS-III BOSS Data Release 12 large-scale structure catalog for the
\texttt{CMASSLOWZTOT} galaxy sample\footnote{\url{https://data.sdss.org/sas/dr12/boss/lss/}} \citep{Reid:2016}.
Each BOSS galaxy is assigned a combined completeness weight, $w_c$, to correct for redshift failures,
fiber collisions
and imaging systematics as
$
w_{c}=w_{\rm sys}(w_{\rm fc} + w_{\rm rf} - 1)
$.
We further weight each galaxy at redshift $z$ by
the FKP weight $w_{\mathrm{FKP,g}}(z)$ \cite{Feldman:1994}.
Hereafter, we call this sample the ``density sample'' and use it to measure the galaxy density field.
For the random catalog, we use the \texttt{random0} file corresponding to the \texttt{CMASSLOWZTOT}
sample, which contains 50 times as many random points as galaxies.

Galaxy ellipticities are defined by two components,
\begin{align}
\gamma_{(+,\times)}=\frac{1-q^2}{1+q^2} 
\left( \cos{(2\beta)}, \sin{(2\beta)}\right)
\label{eq:ellip}
\end{align}
where $q$ is the minor-to-major-axis ratio ($0\leq q \leq 1$) and
$\beta$ is the position angle of the ellipticity.
To obtain higher-quality shape measurements for the BOSS galaxies,
we follow Ref.~\cite{Xu:2023} and cross-match them with DESI Legacy Imaging Survey DR9 data\footnote{\url{https://www.legacysurvey.org/dr9/files/}} \citep{DESI_Collaboration:2016,Dey:2019}.
The DESI imaging covers the entire BOSS footprint and is two to three magnitudes deeper than the SDSS photometry used for target selection, enabling robust measurements of galaxy orientations.
Motivated by Ref.~\cite{Xu:2023}, we use galaxies with S\'ersic index $n>2$ \cite{Sersic:1963}, which preferentially selects early-type galaxies with strong shape alignments.
Following Ref.~\cite{Kurita:2023}, we apply the same completeness weight $w_c$ to the shape sample. Since the shape field is shot-noise dominated, its FKP weight, $w_{\mathrm{FKP,\gamma}}(z)$, is nearly constant across the redshift range and has a negligible effect on our IA statistics \cite{Kurita:2023}.
Because we marginalize over the IA amplitude and use only the galaxy orientations, we set the axis ratio
to $q=0$, as in our earlier studies
\citep{Okumura:2009a,Okumura:2009,Okumura:2023,Okumura:2025}.
We refer to this sample as the ``shape sample.''

%%%%%%%%%%%%%%%%%%%%%%%%%%%%%%%%%%%%%%%%%
% Section 3
%%%%%%%%%%%%%%%%%%%%%%%%%%%%%%%%%%%%%%%%%

%\section{Measurement of correlation functions}\label{sec:measurement}
%{\it Correlation function estimators}---
{\it Measurements of IA correlation functions}---The galaxy position--intrinsic
ellipticity (GI) correlation, $\xigp$, and intrinsic
ellipticity-ellipticity (II) correlations, $\xip$ and $\xim$, are
defined by
\be
\xiX(\bfr) = \left\langle \left[1+\deltag(\bfx_1)\right] \left[1+\deltag(\bfx_2)\right] \WX(\bfx_1,\bfx_2)\right\rangle,
\ee
where ${\rm X}=\{{\rm g}+,{\pm}\}$, $\bfr=\bfx_2-\bfx_1$, $\delta_{\rm g}(\bfx)$ is the galaxy number density fluctuation, $\Wgp(\bfx_1,\bfx_2) = \gamma_+(\bfx_2)$, and
$\Wpm(\bfx_1,\bfx_2) = \gamma_+(\bfx_1)\gamma_+(\bfx_2) \pm
\gamma_\times(\bfx_1)\gamma_\times(\bfx_2)$. 
Since $\xip$ is much noisier than $\xim$ \cite{Okumura:2009,Okumura:2023,Okumura:2025}, 
we do not consider it throughout this paper and refer to $\xim$ simply as the II correlation.

Following Ref.~\cite{Mandelbaum:2006}, we estimate the GI and II correlations as
\begin{align}
&\xigp(\bfr)=\frac{S_+(D-R) }{R_sR}\,, \quad 
\xim(\bfr)=\frac{S_+S_+ - S_\times S_\times}{R_sR_s}\ , \label{eq:estimator}
\end{align}
where $S_+D$ is the sum over all pairs with separation $\bfr$ of the
$+$ component of the ellipticity, $S_+D = \sum_{i\neq j| \bfr}
{\gamma_+(j|i)}$, with $\gamma_+(j|i)$ being the ellipticity of galaxy
$j$ measured relative to the direction to galaxy $i$, 
$S_+S_+ = \sum_{i\neq j| \bfr} {\gamma_+(j|i)\gamma_+(i|j)}$,
and $S_+R$ and $S_\times S_\times$ are defined analogously to $S_+D$ and $S_+S_+$, respectively.
The denominators $R_sR$ and $R_sR_s$ are the normalized shape--density and shape--shape random pair counts, respectively.

To analyze anisotropic features of the IA correlations, we bin galaxy pairs in both radial separation $r = |\bfr |$ and the cosine of the angle to the line of sight, $\mu_{\bfr}$. 
The GI and II correlation functions are naturally expanded in terms of the associated Legendre polynomials ${\cal L}^m_\ell$, with $m=2$ and $4$, respectively
\cite{Kurita:2022,Okumura:2024,Singh:2024,Inoue:2025,Okumura:2025}:
\begin{align}
\xitXlm
(r) &=2\sum_{i}\Delta\mu_{\bfr} \,  \xiX (r,\mu_{\bfr,i})\, \Theta_{\ell}^{m}(\mu_{\bfr,i})~,
\label{eq:xiX_multipole_associated_legendre}   
\end{align}
where $\Theta_{\ell}^{m}(\mu)$ is the normalized associated Legendre function related to the unnormalized one by
$\Theta_{\ell}^{m}(\mu) =\sqrt{\frac{2\ell+1}{2}\frac{(\ell-m)!}{(\ell+m)!}}\mathcal{L}_{\ell}^{m}(\mu)$.
We use a tilde to distinguish these coefficients from those expanded in the standard Legendre basis. 
The summation in Eq.~(\ref{eq:xiX_multipole_associated_legendre})
is performed over  $0\leq \mu_{\bfr}\leq 1$.

%%%%%%%%%%%%%%%%%%%%%%%%%%%%%%%%%%%%%%%%%
% Figure 1
%%%%%%%%%%%%%%%%%%%%%%%%%%%%%%%%%%%%%%%%%

\begin{figure}[tb]
\begin{center}
\includegraphics[width=\columnwidth]{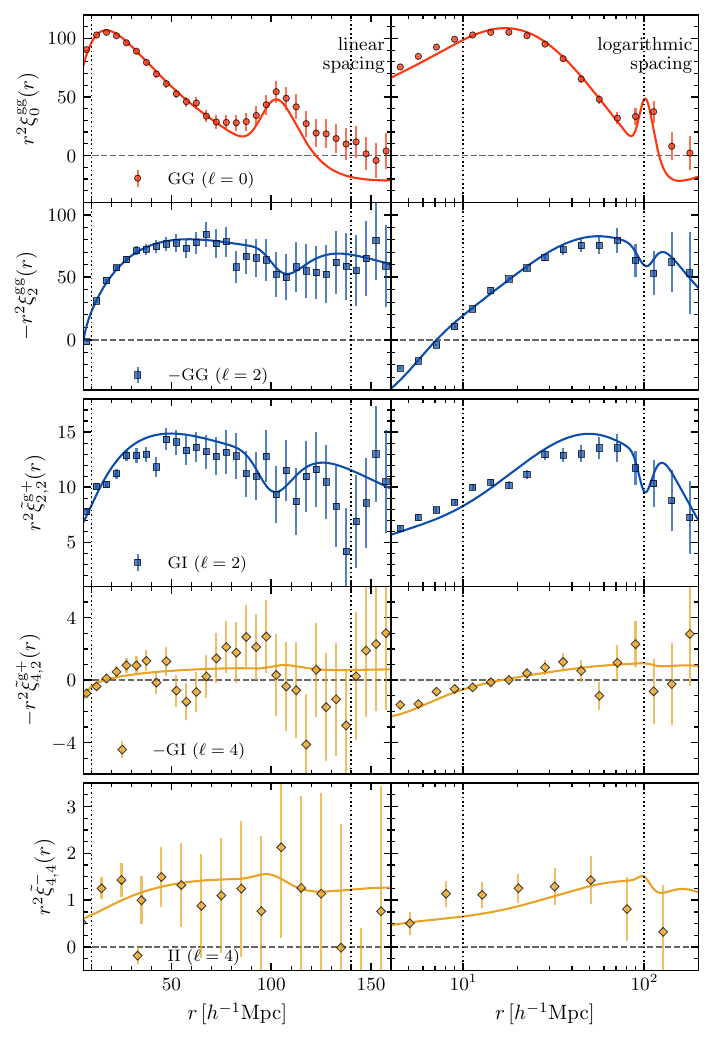}
\caption{
Redshift-space multipole correlation functions of SDSS galaxies with DESI imaging at $0.43<z<0.7$.
The first and second rows show the galaxy autocorrelation monopole and quadrupole, respectively. The third and fourth rows show the GI quadrupole and hexadecapole with $m=2$, and the bottom row shows the leading II multipole, the hexadecapole with $m=4$. Each multipole is rescaled by the power of $r$ indicated on the vertical axis.
The left and right columns show the same measurements with linear and logarithmic radial binning, respectively.
Points with error bars show the measurements, and solid curves show the best-fit models.
Vertical dotted lines show the fitting ranges used in the corresponding cosmological analyses.
}
\label{fig:ximu_gg_gi_ii}
\end{center}
\end{figure}

We use the galaxy autocorrelation (GG) function in redshift space, $\xigg(\bfr)=\langle \deltag(\bfx_1)\deltag(\bfx_2)\rangle$, as our conventional clustering observable.
We adopt the Landy-Szalay estimator \cite{Landy:1993} to measure it.
We obtain the GG multipoles by a standard Legendre expansion and retain
$\xigg_0(r)$ and $\xigg_2(r)$ in the analysis.

Figure~\ref{fig:ximu_gg_gi_ii} presents the measured redshift-space
multipoles of the GG, GI, and II correlation functions for the galaxy
sample at $0.43<z<0.7$.
The left and right columns show the same measurements with linear and
logarithmic radial binning, respectively.
We use the linear-binned measurements for our main joint geometric and
dynamical analysis, since this binning resolves the acoustic feature on
large scales.
The logarithmic-binned measurements instead show the broadband shape
more clearly and are used in Appendix~\ref{app:rsd_only} for the RSD-only
comparison with Ref.~\cite{Okumura:2023}.
The first two rows show the monopole and quadrupole moments of the GG
correlation function.
These statistics have been measured in previous BOSS analyses
\citep{Ross:2017,Satpathy:2017,Chuang:2017}, and our measurements are consistent
with them.

The lower three rows show the IA statistics expanded in the associated
Legendre basis.
The leading GI multipole, $\wt{\xi}^{{\rm g}+}_{2,2}$, is clearly detected over a
wide range of scales, while the $\ell=4$ GI multipole is noisier and
consistent with zero within the present uncertainties.
The leading II multipole, $\wt{\xi}^{-}_{4,4}$, is also detected at lower
significance.

We estimate the covariance matrix for all measured correlation
functions, $\xiggl$, $\xitgplm$, and $\xitmlm$, 
using the jackknife resampling method.
Although the jackknife covariance is not unbiased, it provides
reliable error bars when shot noise dominates \citep{Mandelbaum:2006,Okumura:2025}.
The error bars shown in
Fig.~\ref{fig:ximu_gg_gi_ii} are the square roots of the diagonal
elements of the covariance matrix.

%%%%%%%%%%%%%%%%%%%%%%%%%%%%%%%%%%%%%%%%%
% Section 4
%%%%%%%%%%%%%%%%%%%%%%%%%%%%%%%%%%%%%%%%%

%\section{Theoretical prediction} \label{sec:theory}
{\it Modeling correlation functions}---
Here we present theoretical models to interpret the measured
correlation functions.
The GG model follows the standard nonlinear RSD treatment, while the
new ingredient of this work is the application of the
AP effect to nonlinear IA statistics.
We first present models for the power spectra as functions of the wave number $k$ and directional cosine $\mu_{\bfk}$, $\PX(k,\mu_{\bfk})$, and then transform
them to the correlation functions, $\xiX$.
We adopt the phenomenological RSD model developed for the GG power spectrum in Ref.~\cite{Scoccimarro:2004} and extended to IA power spectra in Ref.~\cite{Okumura:2024}.
The Finger-of-God (FoG) effect is represented by a damping function $D_{\rm FoG}$ and factorized as
\begin{align}
\PX(k,\mu_{\bfk}) = D_{\rm FoG}^2(k\mu_{\bfk}\sigv)\PhX(k,\mu_{\bfk}). \label{eq:pheno_RSD}
\end{align}
We adopt a Gaussian form, $D_{\rm FoG}(k\mu_{\bfk}\sigv)=\exp{\left(-k^2\mu_{\bfk}^2\sigv^2/2\right)}$, characterized by the nonlinear velocity-dispersion parameter $\sigv$.

For the galaxy power spectrum, with ${\rm X=gg}$ in Eq.~(\ref{eq:pheno_RSD}), we adopt the nonlinear RSD model
\citep{Scoccimarro:2004},
\begin{align}
\Phgg(k,\mu_{\bfk}) = b^2 \Pdd(k)   + 2bf\mu_{\bfk}^2 \Pdt(k) + f^2\mu_{\bfk}^4 \Ptt(k), \label{eq:scoccimarro}
\end{align}
where $b$ is the galaxy bias and $f$ is the growth rate. 
We use the revised \texttt{Halofit} prescription to compute the nonlinear density power spectrum $\Pdd$ \citep{Takahashi:2012}. The velocity autopower spectrum $\Ptt$ and density--velocity cross-power spectrum $\Pdt$ are computed with the fitting formulae of Ref.~\cite{Hahn:2015}.
Since $\Pdd$, $\Pdt$, and $\Ptt$ are proportional to the square of the normalization
parameter of the density fluctuation, $\sigma_8^2(z)$, this model contains three free parameters, $\bftheta=(b\sigma_8, f\sigma_8,\sigv)$.

To model the IA statistics, we adopt the nonlinear alignment (NLA) model,
which assumes a linear relation between the intrinsic ellipticity and the underlying nonlinear tidal field,
$\left(\nabla^{-2}\nabla_i\nabla_j\right)\delta_m(\bfx) $ 
\citep{Catelan:2001,Hirata:2004,Bridle:2007}.
In Fourier space, the ellipticity projected along the line of sight ($z$-axis) is given by
\begin{align}
\gamma_{(+,\times)}(\bfk)=\bK 
k^{-2}
\left( k_{x}^{2}-k_{y}^{2}, \ 2k_{x}k_{y}  \right)
\delta_m(\bfk),
\end{align}
where $\bK$ is the redshift-dependent IA coefficient,
which we refer to as the shape bias.
The redshift-space shape field is further multiplied by the FoG damping function.
We then define E-/B-modes, $\gamma_{(E,B)}$
as \citep{Crittenden:2002},
\begin{align}
\gammaE(\bfk) + i\gammaB(\bfk) = e^{-2i\phi_k}\left[  \gamma_{+}(\bfk)+i\gamma_{\times}(\bfk) \right],
\label{eq:gamma_E_deltam}
\end{align}
where $\phi_k$ is the azimuthal angle of the wavevector projected on
the celestial sphere.
Adopting the nonlinear RSD model, the gE and EE power spectra in Eq.~(\ref{eq:pheno_RSD}) are expressed as \citep{Okumura:2024}
\begin{align}
\PhgE(k,\mu_{\bfk}) &= \bK(1-\mu_{\bfk}^2)\left[b \Pdd(k) +f\mu_{\bfk}^2\Pdt(k)\right]~, \label{eq:nla_rsd_gi}  \\
\PhEE(k,\mu_{\bfk}) &= \bK^2 (1-\mu_{\bfk}^2)^2 \Pdd(k)~, \label{eq:nla_rsd_ii} 
\end{align}
The FoG effect enters the IA power spectra in the same way as the GG spectrum.

We now apply the AP effect to the three spectra, $\PX$ with ${\rm X=\{gg,gE,EE\}}$, in a
unified way.
For GG this is the standard treatment, while for the IA spectra it
allows us to model geometric distortions over scales encompassing the BAO feature.
Defining $\hH$ and $\hDA$ as $\hH(z)=H(z)/\Hf(z)$, $\hDA(z)=\DA(z)/\DAf(z)$,
the observed power spectra in comoving space are modeled as
\citep{Alcock:1979,Ballinger:1996,Matsubara:1996,Padmanabhan:2008,Taruya:2011}:
\be
\PXobs(k,\mu_{\bfk})=\frac{\hH(z)}{\hDA^{2}(z)}\PX(q,\nu_{\bfq}),
\ee
where 
$q(k,\mu_{\bfk})=\epsilon(\mu_{\bfk})k$ and 
$\nu_{\bfq}(k,\mu_{\bfk})=\hH\epsilon^{-1}(\mu_{\bfk})\mu_{\bfk}$ 
with 
$\epsilon^2(\mu_{\bfk})=\hDA^{-2} + ( \hH^2- \hDA^{-2} ) \mu_{\bfk}^2$.

Multipole moments of the galaxy density power spectrum, $\Pgglobs(k)$, are obtained by numerically integrating over $\mu_{\bfk}$ with the standard Legendre polynomials.
The correlation-function multipoles are obtained by a Hankel transform, $\xigglobs(r) =\mathcal{H}_\ell^{-1}\left[\Pgglobs(k) \right](r)$,
where the inverse Hankel transform of a function $g(k)$ is expressed as
$\mathcal{H}_\ell^{-1}\left[ g(k)\right](r) \equiv i^\ell \int \frac{k^2 dk}{2\pi^2}j_\ell(kr)g(k)$.

Multipole moments of the galaxy IA power spectra can be computed similarly, but using associated Legendre polynomials
\citep{Kurita:2022,Okumura:2024},
\begin{align}
\PtXlmobs(k)&=\int^1_{-1}d\mu_{\bfk}\PXobs(k,\mu_{\bfk})\Theta_{\ell}^{m}(\mu_{\bfk})~.
\end{align}
Finally, the correlation function multipoles are obtained by a Hankel transform,
\begin{align}
\xitgplmobs(r) &=\mathcal{H}_\ell^{-1}\left[\PtgElmobs(k) \right](r),\\
\xitmlmobs(r) &=\mathcal{H}_\ell^{-1}\left[\PtEElmobs(k) \right](r),
\label{eq:xiX_hankel}
\end{align}
where we choose $m=2$ and 4 for the GI and II correlation functions, respectively.

%%%%%%%%%%%%%%%%%%%%%%%%%%%%%%%%%%%%%%%%%
% Section 5
%%%%%%%%%%%%%%%%%%%%%%%%%%%%%%%%%%%%%%%%%

%\section{Constraints on growth rate}
{\it Constraints on growth and expansion rates}---
In our previous work \cite{Okumura:2023}, we constrained the growth-rate
parameter $f\sigma_8$ from the RSD analysis of the SDSS galaxy samples.
Here we extend that analysis and
jointly constrain $f\sigma_8$, $H$, and $\DA$ through RSD and AP distortions over scales encompassing the BAO feature.
For the clustering-only analysis, our data vector consists of
$\xigg_0$ and $\xigg_2$; for the GG+IA analysis, we additionally include $\xitgp_{2,2}$, $\xitgp_{4,2}$, and $\xitm_{4,4}$.
We use the full covariance matrix to account for
correlations among separation bins and among
different statistics, although the IA correlations are dominated
by shot noise and thus primarily by the diagonal elements.
We have five free parameters for the clustering-only analysis,
${\bftheta}=(f\sigma_8,H,\DA,b\sigma_8,\sigv)$, and include the additional parameter $\bK\sigma_8$ in the GG+IA analysis.
Because we set $q=0$ in Eq.~(\ref{eq:ellip}), our normalization of
$\bK$ differs from conventional definitions in the literature, and its
value should not be compared directly with published measurements.
We perform the analysis over $\rmin \leq r_i \leq \rmax$.
Because jackknife estimates tend to underestimate the covariance on
large scales, we set $\rmax=140\mpcoh$ to limit this effect while retaining the BAO information.
As our fiducial choice, we adopt $\rmin=10\mpcoh$ for all the
statistics, which avoids scales
on which the nonlinear modeling is expected to be least reliable.
The data points used for the analysis are enclosed by the
vertical lines in the left column of Fig.~\ref{fig:ximu_gg_gi_ii}.
For the clustering-only analysis, the covariance is
a $52\times52$ matrix, while for the full analysis of clustering and
IA, it is a $117\times117$ matrix. 

%%%%%%%%%%%%%%%%%%%%%%%%%%%%%%%%%%%%%%%%%
% Figure 2
%%%%%%%%%%%%%%%%%%%%%%%%%%%%%%%%%%%%%%%%%

\begin{figure}[t]
\begin{center}
\includegraphics[width=\columnwidth]{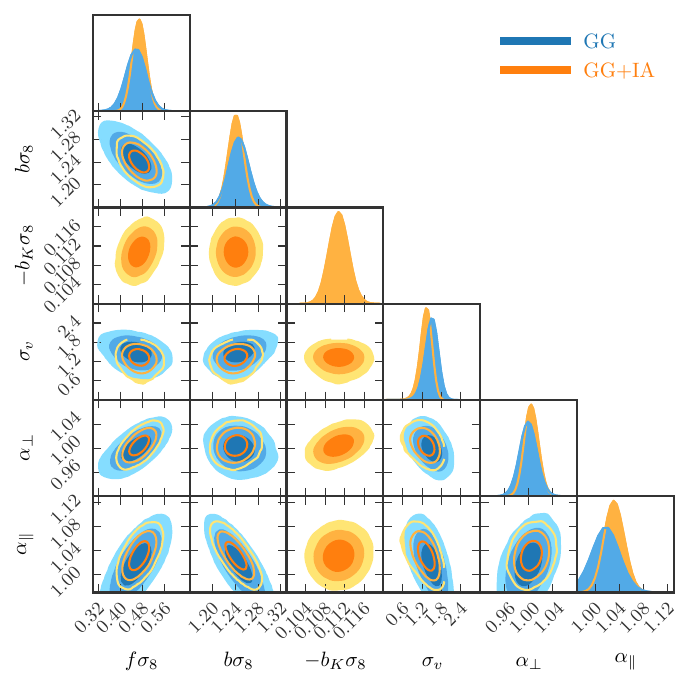}
\caption{Constraints on $(f\sigma_8,\alpha_\perp,\alpha_\parallel,b\sigma_8,\bK\sigma_8,\sigma_v)$, where $\alpha_\perp = \DA/\DAf$ and $\alpha_\parallel = H/\Hf$.
Blue contours show the clustering-only result, while orange contours show the result after adding the IA statistics (GI+II).
The constraints are obtained for the galaxy sample at $0.43<z<0.7$ with $10\leq r \leq 140\mpcoh$.
The contours denote the $68\%$, $95\%$, and $99\%$ credible regions from innermost to outermost.}
\label{fig:result_f_DA_H}
\end{center}
\end{figure}

\begin{table}[tb]
\caption{
Marginalized $68\%$ constraints for the GG-only and GG+IA analyses, where IA denotes the combined GI and II statistics.
For each parameter $\theta$, the improvement is defined as
$1-(\sigma_\theta/|\theta|)_{\rm GG+IA}/
(\sigma_\theta/|\theta|)_{\rm GG}$, where $\sigma_\theta$ is the mean of
the upper and lower marginalized errors.
}
\label{tab:constraints}
\setlength{\tabcolsep}{9pt}
\begin{ruledtabular}
\begin{tabular}{lccc}
$\theta$ & GG & GG+IA & Improvement \\
\hline
\multicolumn{4}{c}{RSD+AP} \\
$f\sigma_8$      & $0.456^{+0.034}_{-0.035}$ & $0.468^{+0.024}_{-0.024}$ & $32\%$ \\
$\alpha_\perp$   & $0.999^{+0.014}_{-0.013}$ & $1.004^{+0.011}_{-0.011}$ & $18\%$ \\
$\alpha_\parallel$ & $1.017^{+0.022}_{-0.022}$ & $1.031^{+0.016}_{-0.016}$ & $29\%$ \\
\hline
\multicolumn{4}{c}{Flat $w_0$CDM} \\
$w_0$            & $-0.815^{+0.080}_{-0.077}$ & $-0.747^{+0.069}_{-0.067}$ & $5\%$ \\
$\Omega_m$       & $0.328^{+0.094}_{-0.076}$ & $0.363^{+0.067}_{-0.058}$ & $33\%$ \\
$H_0$            & $64.8^{+3.0}_{-2.8}$ & $63.3^{+2.1}_{-2.0}$ & $27\%$ \\
\hline
\multicolumn{4}{c}{RSD only} \\
$f\sigma_8$      & $0.463^{+0.015}_{-0.015}$ & $0.459^{+0.014}_{-0.014}$ & $6\%$ \\
\end{tabular}
\end{ruledtabular}
\end{table}

Figure~\ref{fig:result_f_DA_H} shows the parameter constraints
obtained from our galaxy sample.
The blue and orange contours show the clustering-only and GG+IA
results, respectively. The corresponding marginalized constraints
and fractional improvements from adding IA are summarized in
Table~\ref{tab:constraints}.
After marginalizing over the nuisance parameters, the clustering-only
analysis yields $f\sigma_8=0.456^{+0.034}_{-0.035}$,
$\alpha_\perp=0.999^{+0.014}_{-0.013}$, and
$\alpha_\parallel=1.017^{+0.022}_{-0.022}$.  Adding the GI and II
statistics gives $f\sigma_8=0.468^{+0.024}_{-0.024}$,
$\alpha_\perp=1.004^{+0.011}_{-0.011}$, and
$\alpha_\parallel=1.031^{+0.016}_{-0.016}$. The AP parameters are
consistent with their fiducial values, $\alpha_\perp=\alpha_\parallel=1$,
in both analyses. The fractional uncertainties on $f\sigma_8$,
$\alpha_\perp$, and $\alpha_\parallel$ are reduced by $32\%$, $18\%$,
and $29\%$, respectively, when IA is included.
Importantly, this gain is not reproduced in an RSD-only analysis: when the AP parameters are fixed, adding IA improves the $f\sigma_8$ constraint by only 6\% (Appendix ~\ref{app:rsd_only}). The substantial gain in the joint analysis therefore arises from the geometric information carried by the IA anisotropies.
We investigate the dependence of the model-independent constraints on
$\rmin$ in Appendix~\ref{app:rmin}.
The solid curves in the left column of Fig.~\ref{fig:ximu_gg_gi_ii}
show that the best-fit model reproduces the measured GG and IA multipoles.

%%%%%%%%%%%%%%%%%%%%%%%%%%%%%%%%%%%%%%%%%
% Figure 3
%%%%%%%%%%%%%%%%%%%%%%%%%%%%%%%%%%%%%%%%%

\begin{figure}[tb]
\begin{center}
\includegraphics[width=\columnwidth]{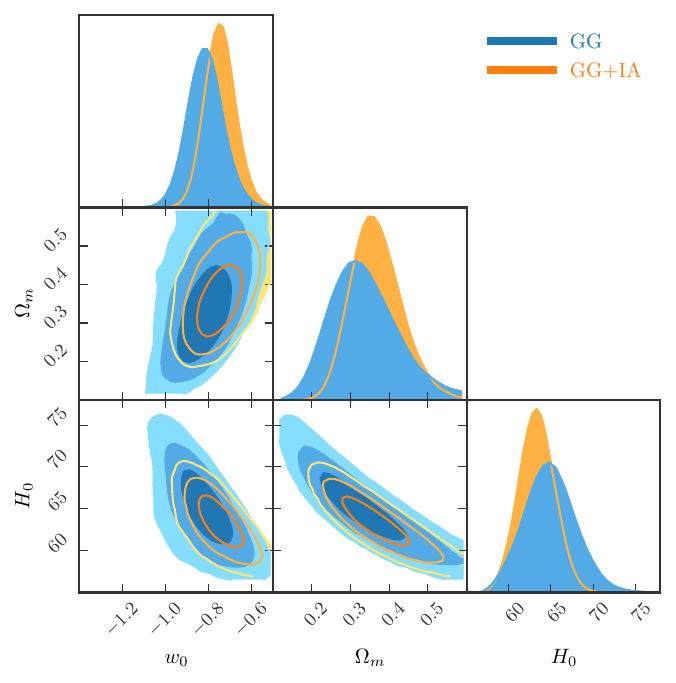}
\caption{
Constraints on the flat $w_0$CDM cosmological model obtained by mapping the model-independent constraints at $0.43<z<0.7$ onto $(w_0,\Omega_m,H_0)$.
The nuisance parameters $b\sigma_8$, $\bK\sigma_8$, and $\sigma_v$ are varied simultaneously.
Blue contours show the GG-only result, while orange contours show the result after adding the IA statistics (GI+II).
The contours denote the $68\%$, $95\%$, and $99\%$ credible regions, and the one-dimensional marginalized posteriors are shown along the diagonal.
}
\label{fig:result_cosmo}
\end{center}
\end{figure}

{\it Constraints on cosmological parameters}---
Finally, we map the model-independent constraints at $0.43<z<0.7$ onto cosmological parameters.
We perform a conditional mapping to a spatially flat $w_0$CDM model in which $(w_0,\Omega_m,H_0)$ are varied, while other cosmological parameters and the broadband power-spectrum shape are held fixed at their fiducial values.
Together with the nuisance parameters, the full parameter vector is
${\bftheta}=(w_0,\Omega_m,H_0,b\sigma_8,\bK\sigma_8,\sigma_v)$.
For each cosmology, we compute $f(z)\sigma_8(z)$, $H(z)$, and $\DA(z)$ and evaluate the corresponding likelihood.

The result is shown in Fig.~\ref{fig:result_cosmo}.
The GG-only constraint exhibits broad degeneracies among $w_0$,
$\Omega_m$, and $H_0$, reflecting the limited information available
from the clustering measurement alone in this single redshift bin.
When the GI and II statistics are included, the allowed parameter
volume is reduced, and the one-dimensional marginalized posteriors
become correspondingly narrower.
This improvement is the cosmological-parameter counterpart of the
tighter constraints on $f\sigma_8$, $\DA$, and $H$ shown in
Fig.~\ref{fig:result_f_DA_H}.
Adding IA breaks part of the clustering-only degeneracy among $w_0,\Omega_m$, and $H_0$, yielding visibly narrower marginalized posteriors.
The marginalized $68\%$ constraints and fractional improvements are summarized
in Table~\ref{tab:constraints}.  The GG-only analysis yields
$w_0=-0.815^{+0.080}_{-0.077}$,
$\Omega_m=0.328^{+0.094}_{-0.076}$, and
$H_0=64.8^{+3.0}_{-2.8}\,{\rm km\,s^{-1}\,Mpc^{-1}}$, while adding IA
gives $w_0=-0.747^{+0.069}_{-0.067}$,
$\Omega_m=0.363^{+0.067}_{-0.058}$, and
$H_0=63.3^{+2.1}_{-2.0}\,{\rm km\,s^{-1}\,Mpc^{-1}}$.
At the fiducial $\rmin=10\mpcoh$, the corresponding improvements are
$5\%$, $33\%$, and $27\%$, respectively. Since the improvement is defined
using the fractional uncertainty, $\sigma_\theta/|\theta|$, the modest
value quoted for $w_0$ partly reflects the shift of its central value
toward smaller $|w_0|$ when IA is included.  In terms of the absolute
uncertainty, the $w_0$ constraint improves by about $13\%$.  The
fiducial GG-only constraint lies about $2.4\sigma$ above $w_0=-1$, and
the addition of IA shifts the central value further toward $w_0>-1$.
However, as shown in Appendix~\ref{app:rmin}, both constraints become
consistent with $w_0=-1$ for $\rmin\geq40\mpcoh$, where the IA improvement in
$w_0$ also becomes negligible.  We therefore do not interpret the
fiducial preference for $w_0>-1$ as evidence against a cosmological
constant; instead, it highlights the sensitivity of the model-dependent
constraint to the small-scale information.

%%%%%%%%%%%%%%%%%%%%%%%%%%%%%%%%%%%%%%%%%
% Section 6
%%%%%%%%%%%%%%%%%%%%%%%%%%%%%%%%%%%%%%%%%

%\section{conclusions} \label{sec:conclusion}
{\it Conclusions}---
We have presented the first joint geometric and dynamical constraints
on cosmology that incorporate the IA of SDSS galaxies measured with
DESI imaging data.  We measured the redshift-space GI and II correlation
functions at $0.43\leq z\leq0.7$ and fitted them with a nonlinear
alignment model that includes RSD and AP distortions of the anisotropic
signal over scales encompassing the BAO feature.  This joint analysis
constrains the growth and expansion rates of the universe through
$f(z)\sigma_8(z)$, $H(z)$, and $\DA(z)$.  Adding IA to galaxy clustering
reduces the fractional uncertainties on $f\sigma_8$, $\alpha_\perp$,
and $\alpha_\parallel$ by $32\%$, $18\%$, and $29\%$, respectively.
Mapping these constraints onto a flat $w_0$CDM model further yields
constraints on $w_0$, $\Omega_m$, and $H_0$; in particular, IA reduces
the absolute marginalized uncertainty on $w_0$ by about $13\%$.  The
preference for $w_0>-1$ obtained with the fiducial scale cut disappears
for $\rmin\geq40\mpcoh$, together with most of the IA improvement in $w_0$.
Thus, the same small-scale information appears to drive both the tighter
constraint and the shift in its central value, and we do not interpret
the latter as evidence against $\Lambda$CDM.

Recent DESI results \cite{DESI:2025a} motivate extending the present analysis to multiple redshift bins and a $w_0w_a$ model. The main limitation is the modeling of nonlinear IA: Fig.~\ref{fig:fsigma8_rmin} shows that the additional constraining power, particularly for $w_0$, depends appreciably on the minimum scale included. Improved nonlinear modeling and simulation-based validation are therefore required before exploiting these scales in precision dark-energy analyses. With such modeling, the same framework can be extended to full-shape IA analyses \cite{Shim:2025,Shim:2025a}, using the broadband matter-power-spectrum information in addition to BAO and RSD \cite{Ivanov:2020,dAmico:2020,Philcox:2020,Kobayashi:2022}.
Such an
analysis will require a sophisticated joint treatment of nonlinear IA, RSD, and
galaxy bias, for which theoretical efforts are actively ongoing
\citep{Blazek:2019,Vlah:2020,Akitsu:2021,Akitsu:2023a,Matsubara:2024,Chen:2024,Taruya:2025,Kurita:2026b}.
Once such modeling is established, the IA contribution can be further
enhanced in future surveys through optimal weighting toward brighter
galaxies \cite{Lamman:2024a,Ishikawa:2025a}.

Ref.~\cite{Xu:2023} applied BAO reconstruction to IA statistics, but
found that the resulting constraint on the isotropic BAO scale was
weaker after reconstruction than before reconstruction. Whether this
degradation arises from the reconstruction implementation or the
response of the IA field requires further study. We therefore do not
apply BAO reconstruction here. Developing and validating a
reconstruction method for anisotropic IA statistics, and determining
whether it sharpens the radial and transverse BAO information, remain
important future work.

Beyond constraints on standard cosmological parameters, galaxy IA has
been proposed as a probe of fundamental physics \citep{Philcox:2024}.
Its spin-dependent correlations are sensitive to primordial
non-Gaussianity and statistical anisotropy
\citep{Schmidt:2015,Shiraishi:2023,Minato:2025}, as well as
primordial gravitational waves, parity violation, and primordial
magnetic fields
\citep{Schmidt:2014,Biagetti:2020,Akitsu:2023,Saga:2024,Okumura:2024a,Kurita:2026,Mikura:2026}.
IA can also test gravity, massive neutrinos,
dark-matter physics, and relativistic effects
\citep{Okumura:2018,Zwetsloot:2022,Chuang:2022,Lee:2023,Saga:2023}.
More broadly, our results show that galaxy shapes need not be treated solely as a systematic of weak-lensing measurements: their anisotropic correlations provide an independent source of geometric and dynamical information that can be exploited alongside galaxy clustering in spectroscopic surveys.

{\it Acknowledgments}---
TO thanks Toshiki Kurita and Atsushi Taruya for discussions
on related projects \cite{Taruya:2025,Kurita:2026b}.
TO acknowledges support from the Taiwan National Science and Technology Council under Grants 
Nos. NSTC 112-2112-M-001-034-,
NSTC 113-2112-M-001-011- and
NSTC 114-2112-M-001-004-, and the Academia Sinica Investigator Project Grant No. AS-IV-114-M03 for the period of 2025-2029.  
Funding for SDSS-III has been provided by the Alfred P. Sloan Foundation, the Participating Institutions, the National Science Foundation, and the U.S. Department of Energy Office of Science. The SDSS-III website is \url{https://www.sdss3.org/}.

%\bibliography{refs}
%\bibliography{ms.bbl}

\appendix

%%%%%%%%%%%%%%%%%%%%%%%%%%%%%%%%%%%%%%%%%
% Figure 4
%%%%%%%%%%%%%%%%%%%%%%%%%%%%%%%%%%%%%%%%%

\begin{figure*}[tb]
\begin{center}
\includegraphics[width = 0.49\textwidth]{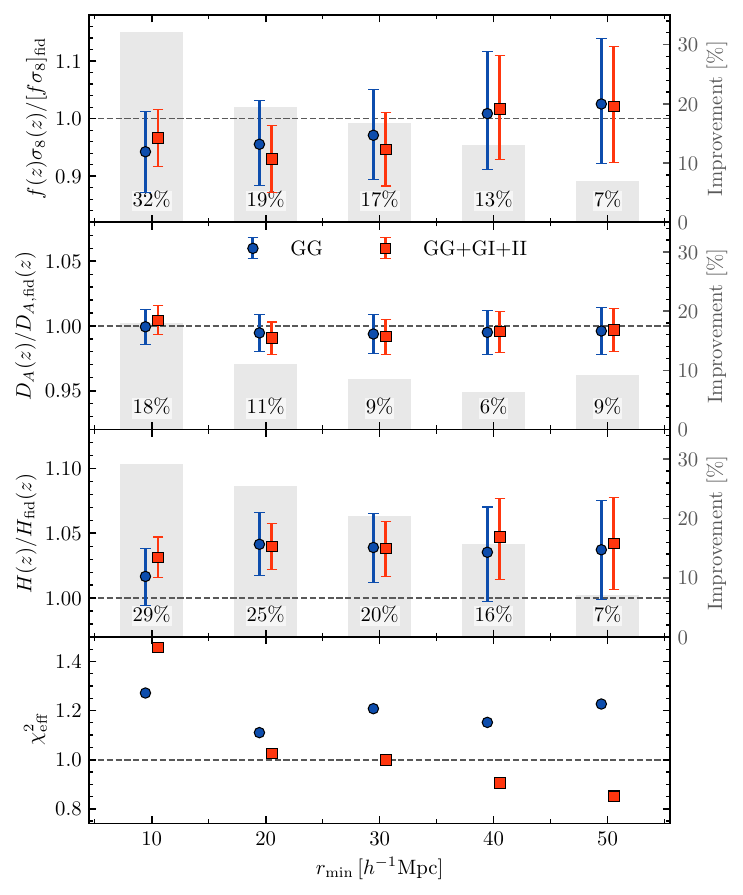}
\includegraphics[width = 0.49\textwidth]{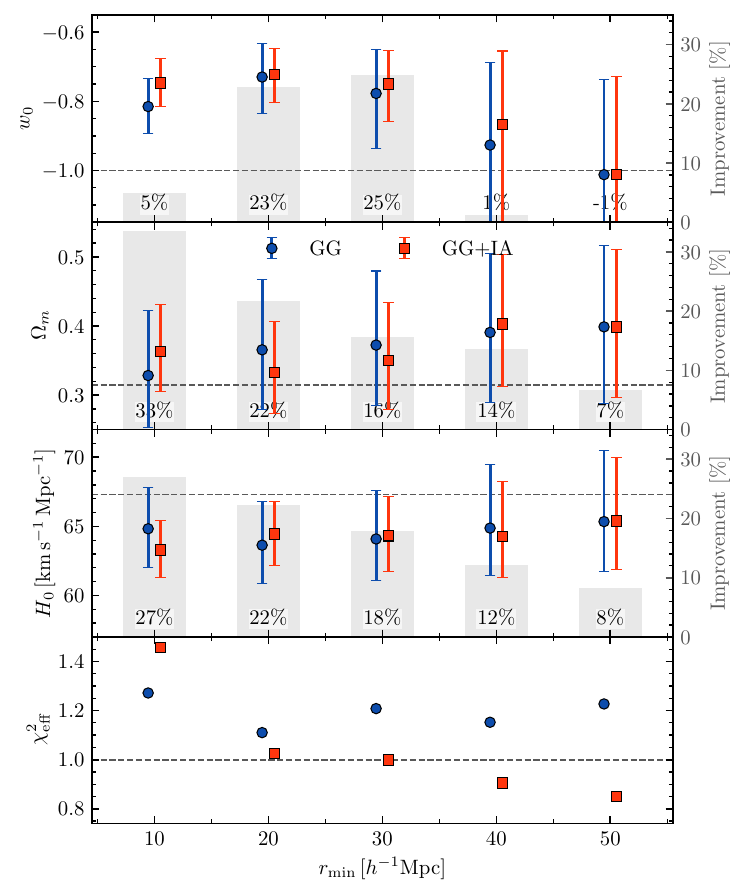}
\caption{
Dependence of the marginalized constraints on the minimum fitting scale, $\rmin$, for the joint geometric and dynamical analysis of the galaxy sample at $0.43<z<0.7$, with $\rmax=140\mpcoh$ fixed.
The upper three panels in the left column show $f(z)\sigma_8(z)/[f\sigma_8]_{\rm fid}$, $\DA(z)/\DAf(z)$, and $H(z)/\Hf(z)$; the corresponding panels in the right column show $w_0$, $\Omega_m$, and $H_0$.
Blue circles use the GG correlation alone, while red squares use the combined GG+GI+II statistics.
The gray bars and right axes show the fractional-error improvement obtained by adding the IA statistics, defined for each parameter $\theta$ as $1-(\sigma_\theta/|\theta|)_{\rm GG+GI+II}/(\sigma_\theta/|\theta|)_{\rm GG}$, where $\sigma_\theta$ is the mean of the upper and lower marginalized $68\%$ errors.
The bottom panels show the effective chi-square, $\chi^2_{\rm eff}$, for the same fits, and the horizontal dashed lines indicate the fiducial parameter values or $\chi^2_{\rm eff}=1$.
}
\label{fig:fsigma8_rmin}
\end{center}
\end{figure*}

%%%%%%%%%%%%%%%%%%%%%%%%%%%%%%%%%%%%%%%%%
% Appendix A
%%%%%%%%%%%%%%%%%%%%%%%%%%%%%%%%%%%%%%%%%

\section{Scale-cut dependence of the joint geometric and dynamical analysis}
\label{app:rmin}

We test the stability of the joint geometric and dynamical constraints
by varying $\rmin$ from $10\mpcoh$ to $50\mpcoh$ while keeping
$\rmax=140\mpcoh$ fixed.  After marginalizing over the nuisance
parameters, Fig.~\ref{fig:fsigma8_rmin} shows the resulting
one-dimensional constraints.  The left column presents the
model-independent constraints on $f\sigma_8$, $\DA$, and $H$, while the
right column presents the corresponding constraints on $w_0$,
$\Omega_m$, and $H_0$ in the flat $w_0$CDM model.

The model-independent constraints remain stable over this range of
scale cuts, with IA consistently tightening their marginalized
uncertainties.  The sensitivity to $\rmin$ becomes more pronounced after
mapping these constraints onto the flat $w_0$CDM model. The improvements in
$\Omega_m$ and $H_0$ are largest at the fiducial scale cut and generally
decrease as smaller scales are removed.  For $w_0$, the fractional-error
improvement is only $5\%$ at the fiducial scale cut because the central value
for GG+IA shifts toward smaller $|w_0|$; the reduction in the absolute
marginalized uncertainty is about $13\%$.  For $\rmin\geq40\mpcoh$, the
constraint broadens and becomes consistent with $w_0=-1$, while the IA
improvement in $w_0$ decreases to nearly zero.

The fact that both the improvement and the shift in $w_0$ are driven by the
small-scale measurements may indicate sensitivity to the accuracy of
the nonlinear IA model, although the present analysis cannot
distinguish a modeling systematic from a statistical fluctuation or
parameter degeneracy.  We therefore regard the $w_0$ scale dependence
as motivation for more detailed modeling and simulation-based
validation of nonlinear IA statistics, rather than as evidence for a
departure from $\Lambda$CDM.  The effective chi-square,
$\chi^2_{\rm eff}=\chi^2_{\rm min}/\nu$, where $\nu$ is the number of
data points minus the number of free parameters, remains of order unity
for both parameterizations.

%%%%%%%%%%%%%%%%%%%%%%%%%%%%%%%%%%%%%%%%%
% Appendix B
%%%%%%%%%%%%%%%%%%%%%%%%%%%%%%%%%%%%%%%%%

\section{RSD-only analysis}
\label{app:rsd_only}
Compared with our previous work
\cite{Okumura:2023}, the present analysis updates both the data and the
methodology: we use higher-quality SDSS galaxy shapes from DESI imaging
and expand the IA statistics in an associated Legendre basis, rather
than the ordinary Legendre basis adopted in Ref.~\cite{Okumura:2023}.
More importantly, the main analysis jointly constrains geometric and
dynamical distortions over a scale range encompassing the BAO feature.
Here we present an RSD-only analysis. Comparing it with the main results
isolates the contribution of the geometric information, while comparing
it with Ref.~\cite{Okumura:2023} reveals the impact of the updated shape
sample and analysis basis.

For the RSD-only analysis, we use the logarithmically binned
correlation functions shown in the right column of
Fig.~\ref{fig:ximu_gg_gi_ii} over the fiducial range
$10\leq r\leq100\mpcoh$.  We analyze the same multipoles as in the
main analysis, $(\xigg_0,\xigg_2,\xitgp_{2,2},\xitgp_{4,2},
\xitm_{4,4})$, but fix the AP parameters to their fiducial values and
vary only $(f\sigma_8,b\sigma_8,\bK\sigma_8,\sigma_v)$.

\begin{figure}[tb]
\begin{center}
\includegraphics[width=\columnwidth]{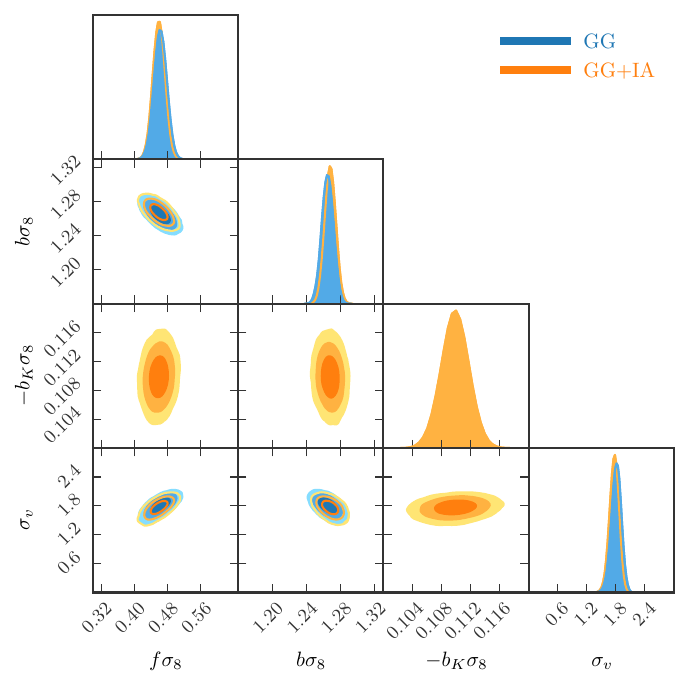}
\caption{
RSD-only constraints on $(f\sigma_8,b\sigma_8,-\bK\sigma_8,\sigma_v)$ for the galaxy sample at $0.43<z<0.7$.
The logarithmically binned correlation functions are used over $10\leq r\leq100\mpcoh$, with the AP parameters fixed to their fiducial values.
Blue contours show the GG-only result, while orange contours show the result after adding the IA statistics (GI+II).
The contours denote the $68\%$, $95\%$, and $99\%$ credible regions, and the one-dimensional marginalized posteriors are shown along the diagonal.
The close overlap of the two results for $f\sigma_8$ shows that IA adds little information in the RSD-only analysis.
}
\label{fig:result_f}
\end{center}
\end{figure}

Figure~\ref{fig:result_f} shows the resulting constraints for the
GG-only analysis and for the combined analysis including the
GI and II statistics.  The constraint on $f\sigma_8$ is consistent with
the CMASS result of Ref.~\cite{Okumura:2023}.
Adding the IA statistics produces almost no improvement in this RSD-only
setup, despite the improved DESI-based shape measurements.
As summarized in Table~\ref{tab:constraints}, the marginalized $68\%$
constraints are $f\sigma_8=0.463\pm0.015$ from GG alone and
$f\sigma_8=0.459\pm0.014$ from GG+IA, corresponding to a fractional-error
improvement of only $6\%$.

\begin{figure}[tb]
\begin{center}
\includegraphics[width=\columnwidth]{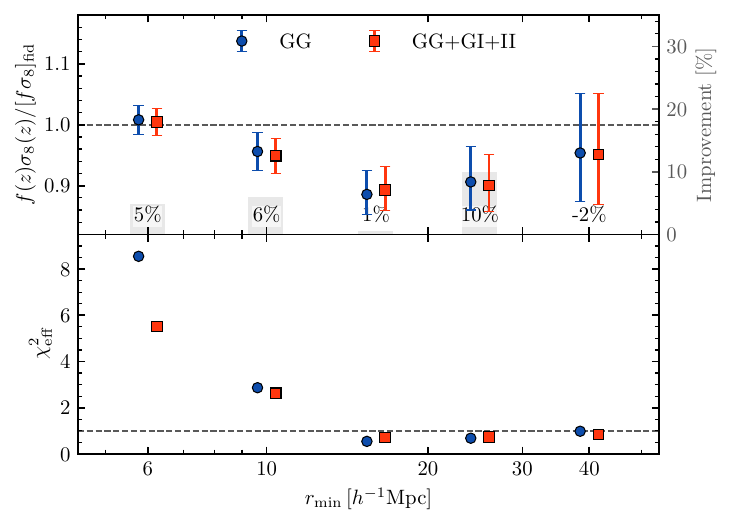}
\caption{
Scale-cut dependence of the RSD-only constraint on $f\sigma_8$ for the galaxy sample at $0.43<z<0.7$, with $\rmax=100\mpcoh$ fixed.
The upper panel shows $f(z)\sigma_8(z)/[f\sigma_8]_{\rm fid}$ as a function of $\rmin$.
Blue circles use the GG correlation alone, while red squares use the combined GG+GI+II statistics.
The gray bars and right axis show the fractional-error improvement obtained by adding the IA statistics, defined as $1-(\sigma_\theta/|\theta|)_{\rm GG+GI+II}/(\sigma_\theta/|\theta|)_{\rm GG}$.
The lower panel shows the effective chi-square, $\chi^2_{\rm eff}$, for the same fits.
}
\label{fig:fsigma8_rmin_log}
\end{center}
\end{figure}

We further test the scale dependence of the RSD-only constraints by
varying $\rmin$ while keeping $\rmax=100\mpcoh$ fixed.
Figure~\ref{fig:fsigma8_rmin_log} shows that the marginalized
$f\sigma_8$ constraints from GG and GG+GI+II remain mutually consistent
over the tested scale cuts.  The fractional improvement from adding IA
fluctuates around zero and remains at only the few-percent level for the
fiducial $\rmin=10\mpcoh$.

%%%%%%%%%%%%%%%%%%%%%%%%%%%%%%%%%%%%%%%%%
% Figure 5
%%%%%%%%%%%%%%%%%%%%%%%%%%%%%%%%%%%%%%%%%

%%%%%%%%%%%%%%%%%%%%%%%%%%%%%%%%%%%%%%%%%
% Figure 6
%%%%%%%%%%%%%%%%%%%%%%%%%%%%%%%%%%%%%%%%%

Thus, the RSD-only analysis does not show a robust cosmological improvement
from IA for the present sample. This should be contrasted with the
main results in Figs.~\ref{fig:result_f_DA_H}--\ref{fig:result_cosmo},
where adding IA significantly improves the cosmological constraints.
The comparison shows that the large improvements in the main analysis
arise from the joint use of IA anisotropies and AP geometric information
over scales encompassing the BAO feature, rather than from the improved
shape catalog alone.

%\clearpage

%\bibliography{refs}
\bibliography{ms.bbl}

%\newpage

\end{document}